\documentclass{llncs}

\usepackage{latexsym,amsfonts,amsmath,amssymb,amscd, graphicx,pgf,tikz}

\renewcommand{\epsilon}{\varepsilon}
\newcommand{\dc}{d_C}

\newcommand{\N}{\mathbb{N}}

\newcommand{\Z}{\mathbb{Z}}
\newcommand{\zz}{{\mathbb{Z}^2}}

\newcommand{\M}{\mathbb{M}}
\newcommand{\U}{\mathbb{U}}

\newcommand{\A}{\mathcal{A}}

\newcommand{\azz}{\mathcal{A}^{\mathbb{Z}^2}}

\newcommand{\am}{\mathcal{A}^{\mathbb{M}}}

\newcommand{\s}{\sigma}

\newcommand{\dd}{\delta}

\newcommand{\e}{\epsilon}

\newcommand{\gs}{\Sigma}

\newcommand\cofin{\textsc{cofin}}
\newcommand\plane[1]{\mathcal{P}_{#1}}

\newcommand{\acf}{F}
\newcommand{\acg}{G}
\newcommand{\acgp}[1]{G_{<#1>}}
\newcommand{\ach}{H}
\newcommand{\aci}{I}
\newcommand{\obst}{\Sigma_{\obstset}}
\newcommand{\nobst}[1]{\Sigma_{\acf,#1}}
\newcommand{\pobst}[1]{\Sigma_{\acg,#1}}
\newcommand{\ppobst}[1]{\Sigma_{<#1>}}
\newcommand{\obstset}{{\mathcal{S}}}
\newcommand{\nobstset}[1]{{\mathcal{R}_{#1}}}
\newcommand{\liquset}{{\mathcal{L}}}
\newcommand{\oT}{\downarrow}
\newcommand{\oB}{\uparrow}
\newcommand{\oL}{\rightarrow}
\newcommand{\oR}{\leftarrow}
\newcommand{\oTL}{\searrow}
\newcommand{\oTR}{\swarrow}
\newcommand{\oBL}{\nearrow}
\newcommand{\oBR}{\nwarrow}
\newcommand{\roto}{\text{$\curvearrowright$}}
\newcommand{\rotb}{\text{\raisebox{1ex}{\rotatebox{180}{$\curvearrowright$}}}}
\newcommand{\slot}[1]{\text{\hbox to 1cm{\hfill\vbox to.5cm{\vfill \hbox{$#1$}\vfill}\hfill}}}
\newcommand{\minislot}[1]{\text{\hbox to.4cm{\hfill\vbox to.2cm{\vfill \hbox{$#1$}\vfill}\hfill}}}
\newcommand{\posout}{\blacksquare}
\newcommand{\posin}{ }
\newcommand{\nslot}[1]{{\hbox to.4cm{\hfill\vbox to.2cm{\vfill \hbox{$#1$}\vfill}\hfill}}}

\newcommand{\nor}[1]{\mathtt{North}\left({#1}\right)}
\newcommand{\sud}[1]{\mathtt{South}\left({#1}\right)}
\newcommand{\est}[1]{\mathtt{East}\left({#1}\right)}
\newcommand{\oue}[1]{\mathtt{West}\left({#1}\right)}
\newcommand{\hau}[1]{\mathtt{Top}\left({#1}\right)}
\newcommand{\bas}[1]{\mathtt{Bot}\left({#1}\right)}
\newcommand{\raisdim}[1]{{#1}^{\uparrow}}
\newcommand{\equpt}{\mathcal{E}_{qu}}
\newcommand{\sensi}{\mathcal{S}_{ens}}
\newcommand{\nono}{\mathcal{N}}

\newcommand{\machine}[1]{\mathcal{M}_{#1}}

\newcommand{\tileset}[1]{\tau_{#1}}

\newcommand{\outo}{\bot}

\newcommand\zplus{+}
\newcommand\zminus{-}
\newcommand\zzero{=}

\newcommand{\transirot}[9]{
  \begin{array}[c]{c}
    \roto{}\\
    \text{$\begin{array}{c|c|c}
          \slot{#1} & \slot{#2} & \slot{#3}\\
          \hline
          \slot{#4} & \slot{#5} & \slot{#6}\\
          \hline
          \slot{#7} & \slot{#8} & \slot{#9}
        \end{array}$}\\
    \rotb{}
  \end{array}\mapsto 
}
\newcommand{\transi}[9]{
    \text{$\begin{array}{c|c|c}
          \slot{#1} & \slot{#2} & \slot{#3}\\
          \hline
          \slot{#4} & \slot{#5} & \slot{#6}\\
          \hline
          \slot{#7} & \slot{#8} & \slot{#9}
        \end{array}$}\mapsto 
}

\title{Topological Dynamics of Cellular Automata: Dimension Matters}

\author{Mathieu Sablik\thanks{\email{sablik@cmi.univ-mrs.fr}}\inst{1}
\and Guillaume Theyssier\thanks{\email{guillaume.theyssier@univ-savoie.fr}}\inst{2}}

\institute{LATP, (UMR 6632 --- CNRS, Universit\'e de Provence), CMI, 
  Universit\'e de Provence, Technop\^ole Ch\^ateau-Gombert, 39, rue F. Joliot Curie, 13453 Marseille Cedex 13 FRANCE 
  \and LAMA, (UMR 5127 --- CNRS, Universit\'e de Savoie), Campus
  Scientifique, 73376 Le Bourget-du-lac cedex FRANCE}
\begin{document}

\maketitle

\begin{abstract}
  Topological dynamics of cellular automata (CA), inherited from
  classical dynamical systems theory, has been essentially studied in
  dimension 1. This paper focuses on higher dimensional CA and aims at
  showing that the situation is different and more complex starting
  from dimension 2. The main results are the existence of non
  sensitive CA without equicontinuous points, the non-recursivity of
  sensitivity constants, the existence of CA having only non-recursive
  equicontinuous points and the existence of CA having only countably
  many equicontinuous points. They all show a difference between
  dimension 1 and higher dimensions.  Thanks to these new
  constructions, we also extend undecidability results concerning
  topological classification previously obtained in the 1D
  case. Finally, we show that the set of sensitive CA is only
  $\Pi_2^0$ in dimension 1, but becomes $\Sigma_3^0$-hard for
  dimension $3$.
\end{abstract}

\section{Introduction}
\nocite{cie}

Cellular automata were introduced by J. von Neumann as a simple formal
model of cellular growth and replication. They consist in a discrete
lattice of finite-state machines, called {\em cells}, which evolve
uniformly and synchronously according to a local rule depending only
on a finite number of neighbouring cells. A snapshot of the states of
the cells at some time of the evolution is called a {\em
  configuration}, and a cellular automaton can be view as a global
action on the set of configurations.

Despite the apparent simplicity of their definition, cellular automata
can have very complex behaviours. One way to try to understand this
complexity is to endow the space of configurations with a topology and
consider cellular automata as classical dynamical systems. With such a
point of view, one can use well-tried tools from dynamical system
theory like the notion of sensitivity to initial condition or the
notion of equicontinuous point.

This approach has been followed essentially in the case of
one-dimensional cellular automata. P.  K{\accent"17 u}rka has shown
in~\cite{Kurka97} that 1D cellular automata are partitioned into two
classes:
\begin{itemize}
\item $\equpt$, the set of cellular automata with equicontinuous points,
\item $\sensi$, the set of sensitive cellular automata.
\end{itemize}
We stress that this partition result is false in general for classical
(continuous) dynamical systems. Thus, it is natural to ask whether
this result holds for the model of CA in any dimension, or if it is a
``miracle'' or an ``anomaly'' of the one-dimensional case due to the
strong constraints on information propagation in this particular
setting.  One of the main contributions of this paper is to show that
this is an anomaly of the 1D case (Section~\ref{sec:core}):
there exist a class $\nono$ of 2D CA which are neither in $\equpt$ nor
in $\sensi$.

Each of the sets $\equpt$ and $\sensi$ has an extremal sub-class:
equicontinous and expansive cellular automata (respectively).  This
allows to classify cellular automata in four classes according to the
degree of sensitivity to initial conditions. The dynamical properties
involved in this classification have been intensively studied in the
literature for 1D cellular automata (see for
instance~\cite{Kurka97,BlanchardMaass,tisseur,permExp}).
Moreover, in~\cite{varouch}, the undecidability of this classification
is proved, except for the expansivity class whose decidability remains
an open problem.

In this paper, we focus on 2D CA and we are particularly interested in
differences from the 1D case. As said above, we will prove in
Section~\ref{sec:core} that there is a fundamental difference
with respect to the topological dynamics classification, but we will
also adopt a computational complexity point of view and show that some
properties or parameters which are computable in 1D are non recursive
in 2D (Proposition~\ref{prop:constant} and \ref{prop:nonrecpoint} of
Section~\ref{sec:classif}). To our knowledge, only few
dimension-sensitive undecidability results are known for CA
(\cite{kari94,Bernardi}).  However, we believe that such subtle
differences are of great importance in a field where the common belief
is that everything interesting is undecidable.

Moreover, we establish in Section~\ref{sec:classif} several complexity
lower bounds on the classes defined above and extend the
undecidability result of~\cite{varouch} to dimension 2. Notably, we
show that each of the class $\equpt$, $\sensi$ and $\nono$ is neither
recursively enumerable, nor co-recursively enumerable. This gives new
examples of ``natural'' properties of CA that are harder than the
classical problems like reversibility, surjectivity or nilpotency
(which are all r.e. or co-r.e.).

Finally, we show two additional results advocating the importance of
dimension in topological dynamics: first, there are 2D CA having only
a countable set of equicontinuous points and, second, the set of
sensitive CA raises from $\Pi_2^0$ in dimension $1$ to
$\Sigma_3^0$-complete in dimension $3$.

\section{Definitions}

Let $\A$ be a finite set and $\M=\Z^d$ (for the $d$-dimensional
case). We consider $\am$, the {\em configuration space} of
$\M$-indexed sequences in $\A$. 

If $\A$ is endowed with the discrete topology, $\am$ is compact,
perfect and totally disconnected in the product topology.  Moreover
one can define a metric on $\am$ compatible with this topology:
$$\forall x,y\in\am,\quad \dc(x,y)=2^{-\min\{\|i\|_\infty: x_i\ne y_i \ i\in\M \}}.$$

Let $\U\subset\M$. For $x\in\am$, denote $x_{\U}\in\A^{\U}$ the
restriction of $x$ to $\U$.  Let $\U\subset\M$ be a finite subset,
$\gs$ is a {\em subshift of finite type of order $\U$} if there exists
$\mathcal{F}\subset\A^{\U}$ such that $x\in\gs \Longleftrightarrow
x_{m+\U}\in\mathcal{F} \quad \forall m\in\M$. In other word, $\gs$ can
be viewed as a tiling where the allowed patterns are in $\mathcal{F}$.

In this paper, we will consider \emph{tile sets} and ask whether they
can tile the plane or not. In our formalism, a tile set is a subshift
of finite type: a set of states (the tiles) given together with a set
of allowed patterns (the tiling constraints).


A {\em cellular automaton} (CA) is a pair $(\am,F)$ where
$F:\am\to\am$ is defined by $F(x)(m)=f((x(m+u))_{u\in\U})$ for all
$x\in\am$ and $m\in\M$ where $\U\subset\Z$ is a finite set named {\em
  neighbourhood} and $f:\A^{\U}\rightarrow\A$ is a {\em local
  rule}. The radius of $F$ is $r(F)=\max\{\|u\|_\infty:u\in\U\}$. By
Hedlund's theorem~\cite{hedlund}, it is equivalent to say that $F$ is
a continuous function which commutes with the shift (i.e. $\s^m\circ
F=F\circ\s^m$ for all $m\in\M$).

We recall here general definitions of topological dynamics used all
along the article. Let $(X,d)$ be a metric space and $F:X\to X$ be a
continuous function.

$\bullet$ $x\in X$ is an {\em equicontinuous point} if for all $\e>0$,
there exists $\dd>0$, such that for all $y\in X$, if $d(x,y)<\dd$ then
$d(F^n(x),F^n(y))<\e$ for all $n\in\N$.

$\bullet$ $(X,F)$ is {\em sensitive} if there exists $\e>0$ such that
for all $\dd>0$ and $x\in X$, there exists $y\in X$ and $n\in\N$ such
that $d(x,y)<\dd$ and $d(F^n(x),F^n(y))>\e$. 

In the definition above about properties of topological dynamics, the
dimension of the cellular automaton considered do not appear
explicitly. Whereas essentially studied in dimension $1$ in the
literature, the present paper consider those properties in any
dimension. A first (trivial) approach to study topological dynamics
properties according to dimension is given by the following
proposition through the notion of canonical lift from dimension $d$ to
dimension $d+1$. The canonical lift of a CA of dimension $d$ with
neighbourhood $\U$ and local rule $f$ is the CA of dimension $d+1$, of
local rule $f$ and of neighbourhood $\U'$ obtained by adding a
coordinate equal to $0$ to each vector of $\U$.

\begin{proposition}
  \label{prop:dimraise}
  Let $F$ be a CA of dimension $d$ and let $\raisdim{F}$ be its
  canonical lift to dimension ${d+1}$. Then we have the
  following:
  \begin{itemize}
  \item $F$ has equicontinuous points if and only if $\raisdim{F}$ has
    equicontinuous points;
  \item $F$ is sensitive to initial conditions if and only if
    $\raisdim{F}$ is sensitive to initial conditions.
  \end{itemize}
\end{proposition}
\begin{proof}
  Straightforward.\qed
\end{proof}

This proposition essentially says that what can be ``seen'' in
dimension $d$ (concerning some topological dynamics properties) can
also be ``seen'' in dimension $d+1$. One of the main point of the
present paper is to show that the converse is false: some behaviours
cannot be ``seen'' in low-dimensional cellular automata.

\section{The Core Construction}
\label{sec:core}

In this section, we will construct a 2D CA which has no equicontinuous
point and is not sensitive to initial conditions. This is in contrast
with dimension 1 where any non-sensitive CA must have equicontinuous
points as shown in~\cite{Kurka97} (such differences according to
dimension will be further discussed in Section~\ref{sec:classif}).

The CA (denoted by $\acf$ in the following) is made of two components:
\begin{itemize}
\item a \emph{solid component} (almost static) for which only
  finite type conditions are checked and corrections are made locally
  ;
\item a \emph{liquid component} whose overall behaviour is to
  infiltrate the solid component and allow some particles to move left
  and to bypass solid obstacles.
\end{itemize}

The general behaviour of this cellular automaton can be seen as an
erosion/infiltration process. States from the solid component can be
turned into liquid state according to certain local conditions but the
converse is impossible. Therefore the set of solid states is
decreasing (erosion process) until some particular kind of
configuration is reached (erosion result). Then, in such
configurations, the particles can bypass any sequence of obstacles and
reach any liquid position (infiltration).

\subsection{Definition}
\label{sec:sketch}

Formally, $\acf$ has a Moore's neighbourhood of radius $2$ ($25$
neighbours) and a state set $\A$ with 12 elements :
${\A=\bigl\{U,D,0,1,\oT,\oB,\oR,\oL,\oTR,\oTL,\oBR,\oBL\bigr\}}$ where
the subset ${\obstset=\{1,\oT,\oB,\oR,\oL,\oTR,\oTL,\oBR,\oBL\}}$
corresponds to the solid component and ${\liquset = \{U,D,0\}}$ to the
liquid component where $0$ should be thought as the substratum where
particles made of elementary constituents $U$ and $D$ can move.

Let $\obst$ be the subshift of finite type of ${\azz}$ defined by the
set of allowed patterns constituted by all the $3\times 3$ patterns
appearing in the following set of finite configurations:

\[
\begin{matrix}
  \minislot{\liquset} & \minislot{\liquset} & \minislot{\liquset} & \minislot{\liquset} & \minislot{\liquset} & \minislot{\liquset} & \minislot{\liquset} & \minislot{\liquset} & \minislot{\liquset} & \minislot{\liquset} \\
  \minislot{\liquset} & \minislot{\liquset} & \minislot{\liquset} & \minislot{\liquset} & \minislot{\liquset} & \minislot{\liquset} & \minislot{\liquset} & \minislot{\liquset} & \minislot{\liquset} & \minislot{\liquset} \\
  \minislot{\liquset} & \minislot{\liquset} & \minislot{\liquset} & \minislot{\oTL} & \minislot{\oT} & \minislot{\oT} & \minislot{\oT} & \minislot{\oTR} & \minislot{\liquset} & \minislot{\liquset} \\
  \minislot{\liquset} & \minislot{\liquset} & \minislot{\liquset} & \minislot{\oL} & \minislot{1} & \minislot{1} & \minislot{1} & \minislot{\oR} & \minislot{\liquset} & \minislot{\liquset} \\
  \minislot{\liquset} & \minislot{\liquset} & \minislot{\liquset} & \minislot{\oL} & \minislot{1} & \minislot{1} & \minislot{1} & \minislot{\oR} & \minislot{\liquset} & \minislot{\liquset} \\
  \minislot{\liquset} & \minislot{\liquset} & \minislot{\liquset} & \minislot{\oL} & \minislot{1} & \minislot{1} & \minislot{1} & \minislot{\oR} & \minislot{\liquset} & \minislot{\liquset} \\
  \minislot{\liquset} & \minislot{\liquset} & \minislot{\liquset} & \minislot{\oBL} & \minislot{\oB} & \minislot{\oB} & \minislot{\oB} & \minislot{\oBR} & \minislot{\liquset} & \minislot{\liquset} \\
  \minislot{\liquset} & \minislot{\liquset} & \minislot{\liquset} & \minislot{\liquset} & \minislot{\liquset} & \minislot{\liquset} & \minislot{\liquset} & \minislot{\liquset} & \minislot{\liquset} & \minislot{\liquset} \\
  \minislot{\liquset} & \minislot{\liquset} & \minislot{\liquset} & \minislot{\liquset} & \minislot{\liquset} & \minislot{\liquset} & \minislot{\liquset} & \minislot{\liquset} & \minislot{\liquset} & \minislot{\liquset}
\end{matrix}
\]
\medskip

Intuitively, $\obst$ defines the 'admissible' solid obstacles,
\textit{i.e.} solid shapes that are stable and no longer eroded in a
liquid environment.

\begin{figure}
  \centering
  \begin{tikzpicture}
    \fill[fill=black] (0,0)--(3,0)--(3,3)--(0,3);
    \draw[very thick, ->] (-.5,.2) node[right] {D} --(-2,.2) node[left] {D};
    \draw[very thick, ->] (-.5,1) node[right] {U} --(-2,1) node[left] {U};
    \draw[very thick, <-] (3.5,2.8) node[left] {U} --(5,2.8) node[right] {U};
    \draw[very thick, <-] (3.5,2) node[left] {D} --(5,2) node[right] {D};
    \draw[very thick, ->] (1,3.5)..controls  (-.6,3.6)..(-.5,2);
    \draw[very thick, ->] (3.5,1)..controls  (3.6,-.6)..(2,-.5);
  \end{tikzpicture}

  \caption{A particle separating into two parts ($U$ and $D$) to
    bypass a solid obstacle (the black region).}
  \label{fig:partdyn}
\end{figure}
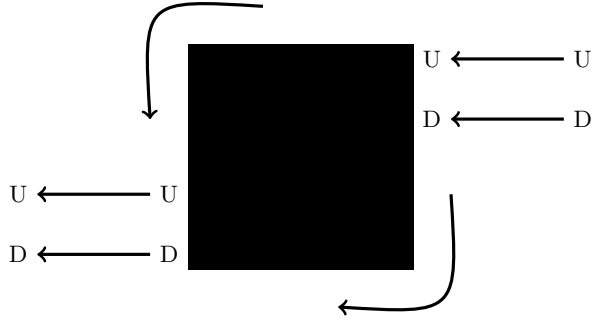

The local transition function of $\acf$ can be sketched as follows:
\begin{itemize}
\item states from $\obstset$ are turned into $0$'s if finite type
  conditions defining $\obst$ are violated locally and left unchanged
  in any other case ;
\item states $U$ and $D$ behave like a left-moving particle when $U$
  is just above $D$ in a background of $0$'s, and they separate to
  bypass solid obstacles, $U$ going over and $D$ going under, until
  they meet at the opposite position and recompose a left-moving
  particle (see Figure~\ref{fig:partdyn}).
\end{itemize}

A precise definition of the local transition function of $\acf$ is the following:
\begin{enumerate}
\item if the neighbourhood ($5\times 5$ cells) forms a pattern
  forbidden in $\obst$, then turn into state $0$ ;
\item else, apply (if possible) one of the transition rules depending
  only on the ${3\times 3}$ neighbourhood detailed in
  Figure~\ref{fig:defi};
  \begin{figure}
    \centering
    {\small\[\begin{array}{c@{\hspace{.5cm}}c@{\hspace{.5cm}}c@{\hspace{.5cm}}c}
        \transirot{\liquset}{\liquset}{\liquset}{\obstset}{x\in\obstset}{\liquset}{\obstset}{\obstset}{\liquset}x,&
        \transirot{\obstset}{\obstset}{\liquset}{\obstset}{x\in\obstset}{\liquset}{\obstset}{\obstset}{\liquset}x,&
        \transi{\obstset}{\obstset}{\obstset}{\obstset}{x\in\obstset}{\obstset}{\obstset}{\obstset}{\obstset}x,\\\\
        \transi{0\text{ or }\obstset}{0}{0}{\obstset}{0}{0}{\obstset}{U}{0}U, &
        \transi{0}{0}{0}{0}{0}{0}{\obstset}{U}{0}U, &
        \transi{0}{0}{0}{0}{0}{U}{\obstset}{\obstset}{D\text{ or }\obstset}U,
        \\\\ \transi{0}{0}{0}{0}{0}{U}{0}{0\text{ or }\obstset}{\obstset}U, &
        \transi{0}{U}{0\text{ or }\obstset}{0}{0}{\obstset}{0}{0}{\obstset}U, & \transi{0}{U}{\obstset}{0}{0}{\obstset}{0}{0}{D}U, \\\\
        \transi{\obstset}{D}{0}{\obstset}{0}{0}{0\text{ or }\obstset}{0}{0}D, &
        \transi{\obstset}{D}{0}{0}{0}{0}{0}{0}{0}D, &
        \transi{\obstset}{\obstset}{U\text{ or }\obstset}{0}{0}{D}{0}{0}{0}D,
        \\\\ \transi{0}{0\text{ or }\obstset}{\obstset}{0}{0}{D}{0}{0}{0}D, &
        \transi{0}{0}{\obstset}{0}{0}{\obstset}{0}{D}{0\text{ or }\obstset}D,& \transi{0}{0}{U}{0}{0}{\obstset}{0}{D}{\obstset}D\\\\
        \transi{0\text{ or }\obstset}{0\text{ or }\obstset}{U}{0\text{ or }\obstset}{0}{D}{0\text{ or }\obstset}{0\text{ or }\obstset}{0\text{ or }\obstset}D,& \transi{0\text{ or }\obstset}{0\text{ or }\obstset}{0\text{ or }\obstset}{0\text{ or }\obstset}{0}{U}{0\text{ or }\obstset}{0\text{ or }\obstset}{D}U\\
      \end{array}\]}
    \caption{Part of the transition rule of $\acf$ (curved arrows
      mean that the transition is the same for any rotation of the
      neighbourhood pattern by an angle multiple of $\pi/2$).}
\label{fig:defi}
\end{figure}

\item in any other case, turn into state $0$.
\end{enumerate}

Note for instance, that any solid state surrounded by a valid
neighbourhood is left unchanged by $\acf$ (second case of the
definition above apply since the $3$ first transitions of
Figure~\ref{fig:defi} include all possible valid ${3\times 3}$
neighbourhoods seen by a solid state).

\subsection{Erosion and Infiltration}
\label{sec:ero}

A configuration $x$ is said to be \emph{finite} if the set ${\bigl\{z
  : x(z)\not=0\bigr\}}$ is finite. The next lemma shows that $\obst$
attracts any finite configuration under the action of
$\acf$. Moreover, after some time, all particles are on the left of
the finite solid part.

\begin{lemma}[erosion process]
  \label{lem:finiteattrak}
  For any finite configuration $x$, there exists $t_0$ such that
  ${\forall t\geq t_0}$ : ${\acf^t(x)\in\obst}$ and, in $\acf^t(x)$,
  any occurrence of $U$ or $D$ is on the left of any occurrence of any
  state from $\obstset$.
\end{lemma}
\begin{proof}
  First, the set ${\bigl\{z : x(z)\in\obstset\}}$ is finite and
  decreasing under the action of $\acf$. Moreover, $U$ and $D$ states
  can only move left, or move vertically or disappear. Since the total
  amount of vertical moves for $U$ and $D$ states is bounded by the
  cardinal of ${\bigl\{z : x(z)\in\obstset\}}$, there is a time $t$
  after which all $U$ or $D$ state are on the left of all occurrences
  of states from $\obstset$, and each $U$ is above a $D$ in a $0$
  background (the $UD$ particle is on the left of the finite non-$0$
  region).  From this time on, the evolution of cells in a state of
  $\obstset$ is governed only by the first case of the definition of
  $\acf$. Therefore, after a certain time, finite type conditions
  defining $\obst$ are verified everywhere. To conclude, it is easy to
  check that $\obst$ is stable under the action of $\acf$.\qed
\end{proof}

The following lemma states that finite configurations from $\obst$
consist of rectangle obstacles inside a liquid background. Moreover,
obstacles are spaced enough to ensure that any position ``sees'' at
most one obstacle in its $3\times 3$ neighbourhood.

In the sequel we use notation $\sud{\cdot}$, $\est{\cdot}$,
$\oue{\cdot}$, $\nor{\cdot}$ for the elementary translations in $\zz$.

\begin{lemma}[erosion result]
  \label{lem:eroded}
  Let ${x\in\obst}$ be a finite configuration. Then the set ${X =
    \{z\in\zz : x(z)\in\obstset\}}$ is a union of disjoint rectangles
  which are pairwise spaced by at least 2 cells.
\end{lemma}
\begin{proof}
  Straightforward from definition of $\obst$.\qed
\end{proof}

An \emph{obstacle} is a (finite) rectangular region of states from
$\obstset$ surrounded by liquid states.

The following lemma establishes the key property of the dynamics of
$\acf$: particles can reach any liquid position inside a finite field
of obstacles from arbitrarily far away from the field.

\begin{lemma}[infiltration]
  \label{lem:partattack}
  Let ${x\in\obst}$ be a finite
  configuration. For any $z_0\in\zz$ such that $x(z_0)=0$ there exists
  a path ${(z_n)}$ such that:
  \begin{enumerate}
  \item ${\|z_n\|_\infty\rightarrow\infty}$
  \item ${\exists n_0,\forall n\geq n_0}$, if $x_n$ is the
    configuration obtained from $x$ by adding a particle at position
    $z_n$ (precisely, ${x_n(z_n)=U}$ and
    ${x_n\bigl(\sud{z_n}\bigr)=D}$) then
    ${\bigl(\acf^n(x_n)\bigr)(z_0)\in\{U,D\}}$.
  \end{enumerate}
\end{lemma}
\begin{proof}
  First, we suppose that
  ${x\in\obst\cap\bigl(\{0\}\cup\obstset\bigr)^\zz}$.  Since
  $x\in\obst$ and $x(z_0)=0$, then either ${x\bigl(\sud{z_0}\bigr)=0}$
  or ${x\bigl(\nor{z_0}\bigr)=0}$ by Lemma~\ref{lem:eroded}. We will
  consider only the first case since the proof for the second one is
  similar. Let $(z_n)$ be the path starting from $z_0$ defined as
  follows:
  \begin{itemize}
  \item If $x\bigl(\est{z_n}\bigr)=0$ and $x\bigl(\sud{\est{z_n}}\bigr)=0$ then ${z_{n+1}=\est{z_n}}$.
  \item Else, position $\est{z_n}$ and/or position $\sud{\est{z_n}}$
    belongs to an obstacle $P$.  Let $a$, $b$ and $c$ be the positions
    of the upper-left, upper-right and lower-right outside corners of
    $P$ and let $p$ be its half perimeter. Then define
    ${z_{n+1},\ldots,z_{n+p+1}}$ to be the sequence of positions made
    of (see Figure~\ref{fig:thepath}):

    \def\blok#1#2{\draw[fill=gray!20!white,shape=square] #1 ++(.5,-.5)-- ++(0,1)-- ++(-1,0)-- ++(0,-1)--cycle;%
      \draw #1 node {#2};}
    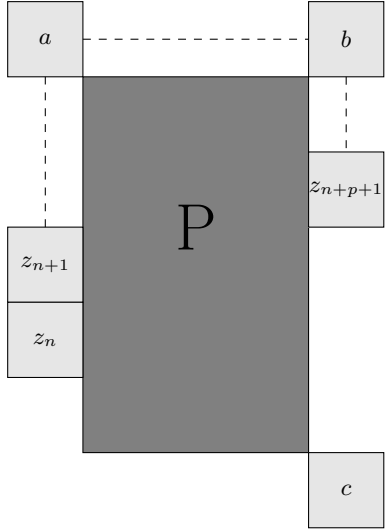
\begin{figure}
      \centering
      \begin{tikzpicture}
        \draw[fill=gray] (0,-1)--(3,-1)--(3,4)--(0,4)--cycle;
        \draw (1.5,2) node {\Huge P};
        \blok{(3.5,4.5)}{$b$}
        \blok{(3.5,-1.5)}{$c$}
        \blok{(-.5,4.5)}{$a$}
        \blok{(-.5,1.5)}{$z_{n+1}$}
        \blok{(3.5,2.5)}{$z_{n+p+1}$}
        \blok{(-.5,0.5)}{$z_{n}$}
        \draw[dashed] (-.5,2)--(-.5,4);
        \draw[dashed] (3.5,4)--(3.5,3);
        \draw[dashed] (0,4.5)--(3,4.5);
      \end{tikzpicture}
      \caption{Definition of the path $(z_n)_n$ in the presence of obstacles.}
      \label{fig:thepath}
    \end{figure}
    \begin{itemize}
    \item a (possibly empty) vertical segment from $z_n$ to $a$,
    \item the segment $[a;b]$,
    \item a (possibly empty) vertical segment from $b$ to $z_{n+p+1}$
      where $z_{n+p+1}$ is the point on $[b;c]$ such that ${z_na +
        bz_{n+p+1}=bc}$.
    \end{itemize}
  \end{itemize}
  We claim that the path $(z_n)$ constructed above has the properties
  of the lemma. Indeed, one can check that for each case of the
  inductive construction of a point $z_m$ from a point $z_n$ we have:
  \begin{itemize}
  \item ${\|z_m\|_\infty>\|z_n\|_\infty}$,
  \item ${\acf^{m-n}(x_m)(z_n)=U}$ and
    ${\acf^{m-n}(x_m)(\sud{z_n})=D}$.
  \end{itemize}

  The lemma is thus proved for
  ${x\in\obst\cap\bigl(\{0\}\cup\obstset\bigr)^\zz}$. It extends to
  any finite ${x\in\obst}$ because in such a configuration
  Lemma~\ref{lem:finiteattrak} ensures that after some time $t_0$ all
  occurrences of $U$ and $D$ are on the left of $z_0$, whereas the
  path constructed above is on the right of $z_0$. More precisely, if
  $x'$ is the configuration obtained from $x$ by replacing any liquid
  state by $0$, and if $(z_n)_n$ is the path constructed for $x'$,
  then the path ${(z_{t_0+n})_n}$ fulfils the requirements of the
  lemma for $x$.\qed
\end{proof}

\subsection{Topological Dynamics Properties}
\label{sec:topdynprop}

The possibility to form arbitrarily large obstacles prevents $\acf$ from
being sensitive to initial conditions.

\begin{proposition}
  \label{prop:nosens}
  $\acf$ is not sensitive to initial conditions.
\end{proposition}
\begin{proof}
  Let ${\epsilon>0}$. Let $c_\epsilon$ be the configuration everywhere
  equal to $0$ except in the square region of side
  ${2\bigl\lceil-\log\epsilon\bigr\rceil}$ around the center where
  there is a valid obstacle.  ${\forall y\in\azz}$, if
  ${d(y,c_\epsilon)\leq\epsilon/4}$ then ${\forall t\geq 0}$,
  ${d\bigl(\acf^t(c_\epsilon),\acf^t(y)\bigr)\leq\epsilon}$ since a
  well-formed obstacle (precisely, a partial configuration that would
  form a valid obstacle when completed by $0$ everywhere) is
  unalterable for $\acf$ provided it is surrounded by states in
  $\liquset$ (see the 3 first transition rules of case 2 in the
  definition of the local rule): this is guarantied for $y$ by the
  condition ${d(y,c_\epsilon)\leq\epsilon/4}$.\qed
\end{proof}

The erosion and infiltration process described above ensures that
particles can circulate everywhere in the liquid part of finite
configurations. This is the key ingredient of the following
proposition.

\begin{proposition}
  \label{prop:noequ}
  $\acf$ has no equicontinuous points.
\end{proposition}
\begin{proof}
  Assume $\acf$ has an equicontinuous point, precisely a point $x$ which
  verifies ${\forall\epsilon>0,\exists\delta : \forall y,
    d(x,y)\leq\delta\Rightarrow \forall t,
    d\bigl(\acf^t(x),\acf^t(y)\bigr)\leq\epsilon}$.

  Suppose that there is $z_0$ such that ${x(z_0)=0}$ and let
  ${\epsilon = 2^{-\|z_0\|_\infty-1}}$. We will show that the
  hypothesis of $x$ being an equicontinuous point is violated for this
  particular choice of $\epsilon$. Consider any ${\delta>0}$ and let
  $y$ be the configuration everywhere equal to $0$ except in the
  central region of radius ${-\log \left\lceil\delta\right\rceil}$
  where it is identical to $x$. Since $y$ is finite, there exists
  $t_0$ such that ${y_+=\acf^{t_0}(y)\in\obst}$ (by
  Lemma~\ref{lem:finiteattrak}).  Moreover,
  Lemma~\ref{lem:finiteattrak} guaranties that for any positive
  integer $t$, ${\acf^t(y_+)(z_0)=x(z_0)=0}$. Applying
  Lemma~\ref{lem:partattack} on $y_+$ and position $z_0$, we get the
  existence of a path ${(z_n)}$ allowing particles placed arbitrarily
  far away from $z_0$ to reach the position $z_0$ after a certain
  time. For any sufficiently large $n$, we construct a configuration
  $y'$ obtained from $y$ by adding a particle at position $z_n$. By
  the property of ${(z_n)}$, we have:
  ${\acf^n(y)(z_0)\not=\acf^n(y')(z_0)}$ and
  therefore ${d\bigl(\acf^n(y),\acf^n(y')\bigr)>\epsilon}$. Since, if
  ${n>-\log \left\lceil\delta\right\rceil}$, both $y$ and $y'$ are in
  the ball of center $x$ and radius $\delta$, we have the desired
  contradiction.

  Assume now that ${\forall z, x(z)\in\obstset}$. There must exist
  some position $z_0$ such that ${x(z_0)\in\obstset\setminus\{1\}}$
  (it is straightforward to check that the uniform configuration
  everywhere equal to 1 is not an equicontinuous point).  It follows
  from the definition of $\obst$ that $z_0$ belongs to a forbidden
  pattern for $\obst$ (any solid state different from $1$ must have a
  liquid state in its neighbourhood). Therefore
  ${\acf(x)(z_0)=0}$ and we can use the reasoning of the
  previous case of this proof on configuration $\acf(x)$.

  Finally, if ${\forall z, x(z)\not=0}$ and ${\exists z_0,
    x(z_0)\in\{U,D\}}$ then necessarily ${F(x)(z_0)=0}$ and the first
  reasoning of the proof can be applied.\qed
\end{proof}

\section{Variations}
\label{sec:var}


\subsection{Adding Wang Tile Constraints}
\label{sec:wang}

The first variation on $\acf$ we consider is to add some tiling
constraints to the solid component. 


More precisely, for any tile set $\tileset{}$, we define a 2D CA
$\acf_{\tileset{}}$ which is identical to $\acf$ except for the
following modifications:
\begin{itemize}
\item the solid state $1$ is replaced by the set $\tileset{}$ so that
  the state set of $\acf_{\tileset{}}$ is
  ${\A_{\tileset{}}=\bigl\{U,D,0,\oT,\oB,\oR,\oL,\oTR,\oTL,\oBR,\oBL\bigr\}\cup\tileset{}}$
  where the solid component is the subset
  ${\obstset_{\tileset{}}=\{\oT,\oB,\oR,\oL,\oTR,\oTL,\oBR,\oBL\}\cup\tileset{}}$
  and the liquid component is also ${\liquset = \{U,D,0\}}$;
\item the sub-shift of 'admissible' obstacles now becomes
  $\nobst{\tileset{}}$ defined by the set of allowed patterns
  constituted by all the $3\times 3$ patterns appearing in the
  following set of finite configurations:

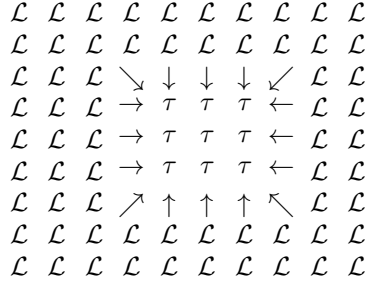
\[
\begin{matrix}
  \minislot{\liquset} & \minislot{\liquset} & \minislot{\liquset} & \minislot{\liquset} & \minislot{\liquset} & \minislot{\liquset} & \minislot{\liquset} & \minislot{\liquset} & \minislot{\liquset} & \minislot{\liquset} \\
  \minislot{\liquset} & \minislot{\liquset} & \minislot{\liquset} & \minislot{\liquset} & \minislot{\liquset} & \minislot{\liquset} & \minislot{\liquset} & \minislot{\liquset} & \minislot{\liquset} & \minislot{\liquset} \\
  \minislot{\liquset} & \minislot{\liquset} & \minislot{\liquset} & \minislot{\oTL} & \minislot{\oT} & \minislot{\oT} & \minislot{\oT} & \minislot{\oTR} & \minislot{\liquset} & \minislot{\liquset} \\
  \minislot{\liquset} & \minislot{\liquset} & \minislot{\liquset} & \minislot{\oL} & \minislot{\tileset{}} & \minislot{\tileset{}} & \minislot{\tileset{}} & \minislot{\oR} & \minislot{\liquset} & \minislot{\liquset} \\
  \minislot{\liquset} & \minislot{\liquset} & \minislot{\liquset} & \minislot{\oL} & \minislot{\tileset{}} & \minislot{\tileset{}} & \minislot{\tileset{}} & \minislot{\oR} & \minislot{\liquset} & \minislot{\liquset} \\
  \minislot{\liquset} & \minislot{\liquset} & \minislot{\liquset} & \minislot{\oL} & \minislot{\tileset{}} & \minislot{\tileset{}} & \minislot{\tileset{}} & \minislot{\oR} & \minislot{\liquset} & \minislot{\liquset} \\
  \minislot{\liquset} & \minislot{\liquset} & \minislot{\liquset} & \minislot{\oBL} & \minislot{\oB} & \minislot{\oB} & \minislot{\oB} & \minislot{\oBR} & \minislot{\liquset} & \minislot{\liquset} \\
  \minislot{\liquset} & \minislot{\liquset} & \minislot{\liquset} & \minislot{\liquset} & \minislot{\liquset} & \minislot{\liquset} & \minislot{\liquset} & \minislot{\liquset} & \minislot{\liquset} & \minislot{\liquset} \\
  \minislot{\liquset} & \minislot{\liquset} & \minislot{\liquset} & \minislot{\liquset} & \minislot{\liquset} & \minislot{\liquset} & \minislot{\liquset} & \minislot{\liquset} & \minislot{\liquset} & \minislot{\liquset}
\end{matrix}
\]
with the additional condition that two adjacent cells in a state from
$\tileset{}$ must fulfils the tiling constraints involved in the tile
set $\tileset{}$.
\end{itemize}

The behaviour of $\acf_{\tileset{}}$ is similar to that of $\acf$
replacing $\obst$ by $\nobst{\tileset{}}$. More precisely:
\begin{enumerate}
\item if the neighbourhood ($5\times 5$ cells) forms a pattern
  forbidden in $\nobst{\tileset{}}$, then turn into state $0$;
\item else, apply (if possible) one of the transition rules depending
  only on the ${3\times 3}$ neighbourhood detailed in
  Figure~\ref{fig:defi} (replacing $\obstset$ by $\obstset_{\tileset{}}$);
\item in any other case, turn into state $0$.
\end{enumerate}

As for $\acf$, the erosion/infiltration mechanism prevents from any
equicontinuous point. Moreover the sensitivity to initial conditions
of $\acf_{\tileset{}}$ is controlled by the tile set $\tileset{}$ as
shown by the following proposition.

\begin{proposition}
  \label{prop:tileac}
  Let $\tileset{}$ be any tile set. Then we have the following:
  \begin{itemize}
  \item $\acf_{\tileset{}}$ has no equicontinuous point;
  \item $\acf_{\tileset{}}$ is sensitive to initial conditions if and
    only if $\tileset{}$ does not tile the plane. Moreover, in this
    case, the maximal sensitivity constant is an exponential function
    of $n$, where ${n\times n}$ is the size of the largest admissible
    square tiling.
  \end{itemize}
\end{proposition}
\begin{proof}
  Firstly, it follows from definition of $\acf_{\tileset{}}$ that
  Lemmas~\ref{lem:finiteattrak}, \ref{lem:eroded} and
  \ref{lem:partattack} as well as Proposition \ref{prop:noequ} remain
  true. Indeed, considering any configuration $x$ of
  $\acf_{\tileset{}}$, and any ${t\geq 0}$, then we have
  \[\{z : \acf_{\tileset{}}^t(x)(z)\in\obstset_{\tileset{}}\} \subseteq
  \{z : \acf^t(x')(z)\in\obstset\}\] where $x'$ is the configuration of
  $\acf$ obtained from $x$ be replacing any occurrence of states from
  $\tileset{}$ by $1$.

  Moreover, if $\tileset{}$ can tile the plane then it is possible to
  form arbitrarily large valid obstacles, so $\acf_{\tileset{}}$ is
  not sensitive to initial conditions (same reasoning as in
  Proposition~\ref{prop:nosens}). Conversely, if $\tileset{}$ cannot
  tile the plane, then there is $n$ such that no valid tiling of a ${
    (2n+1)\times (2n+1)}$ square exists. This implies that, in any
  configuration $x$ of $\acf_{\tileset{}}$, there is some $z_0$ with
  ${\|z_0\|_\infty\leq n}$ such that either $x(z_0)\in\liquset$, or
  ${\acf_{\tileset{}}(x)(z_0)\in\liquset}$ ($z_0$ corresponds to some
  error for $\nobst{\tileset{}}$).  Then, applying
  Lemma~\ref{lem:partattack} to position $z_0$ as in the proof of
  Proposition~\ref{prop:noequ}, we have:
  \[\forall\delta>0, \exists y, \exists t\geq 0 :
  d(x,y)\leq\delta\text{ and
  }d\bigl(\acf_{\tileset{}}^t(x),\acf_{\tileset{}}^t(y)\bigr)\geq 2^{-n}.\]
  Since the constant $n$ is independent of the choice of the initial
  configuration $x$, we have shown that $\acf_{\tileset{}}$ is
  sensitive to initial conditions with sensitivity constant $2^{-n}$.\qed
\end{proof}

\subsection{Controlling Erosion}
\label{sec:onion}

In this section, we define $\acg_{\tileset{}}$, another variant of
$\acf$, which has an overall similar behaviour but uses a different
kind of obstacles and a different kind of erosion process depending on
a tile set $\tileset{}$. Obstacles are protected from liquid component
by a boundary as the classical obstacles of $\acf$, but they are made
only of successive boundaries like onion skins. Moreover, invalid
patterns in the solid component do not provoke the complete
destruction of obstacles as in $\acf$.

The solid component of $\acg_{\tileset{}}$ is the set
${\nobstset{\tileset{}}=\tileset{}\times X}$ where
\[X=\{\oT,\oB,\oR,\oL,\oTR,\oTL,\oBR,\oBL,\outo\}.\] The liquid
component is identical to that of $\acf$, precisely
$\liquset=\{U,D,0\}$.

\begin{figure}
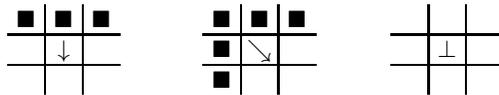

  \centering{
    \begin{tabular}{c|c|c}
      \nslot{\posout} & \nslot{\posout} & \nslot{\posout}\\
      \hline
      \nslot{\posin} & \nslot{\oT} & \nslot{\posin}\\
      \hline
      \nslot{\posin} & \nslot{\posin} & \nslot{\posin}
    \end{tabular} \hskip 1cm
    \begin{tabular}{c|c|c}
      \nslot{\posout} & \nslot{\posout} & \nslot{\posout}\\
      \hline
      \nslot{\posout} & \nslot{\oTL} & \nslot{\posin}\\
      \hline
      \nslot{\posout} & \nslot{\posin} & \nslot{\posin}
    \end{tabular}\hskip 1cm
    \begin{tabular}{c|c|c}
      \nslot{\posin} & \nslot{\posin} & \nslot{\posin}\\
      \hline
      \nslot{\posin} & \nslot{\outo} & \nslot{\posin}\\
      \hline
      \nslot{\posin} & \nslot{\posin} & \nslot{\posin}
    \end{tabular}}
  \caption{Inside (white) and outside (black) positions for states
    of $X$.}
  \label{fig:inout}
\end{figure}

The obstacle sub-shift $\pobst{\tileset{}}$ of $\acg_{\tileset{}}$ is
defined by the set of allowed patterns constituted by all $3\times 3$
patterns appearing in the following set of partial configurations:
\[
\begin{matrix}
  \minislot{\liquset} &   \minislot{\liquset} & \minislot{\liquset} & \minislot{\liquset} & \minislot{\liquset} & \minislot{\liquset} & \minislot{\liquset}& \minislot{\liquset}\\
    \minislot{\liquset} & \minislot{\liquset} & \minislot{\liquset} & \minislot{\nobstset{\tileset{}}} & \minislot{\nobstset{\tileset{}}} & \minislot{\nobstset{\tileset{}}} & \minislot{\liquset} & \minislot{\liquset}\\
    \minislot{\liquset} & \minislot{\liquset} & \minislot{\liquset} & \minislot{\nobstset{\tileset{}}} & \minislot{\nobstset{\tileset{}}} & \minislot{\nobstset{\tileset{}}} & \minislot{\liquset} & \minislot{\liquset}\\
   \minislot{\liquset} & \minislot{\liquset} & \minislot{\liquset} & \minislot{\nobstset{\tileset{}}} & \minislot{\nobstset{\tileset{}}} & \minislot{\nobstset{\tileset{}}} & \minislot{\liquset} & \minislot{\liquset}\\
   \minislot{\liquset} & \minislot{\liquset} & \minislot{\liquset} &  \minislot{\liquset} & \minislot{\liquset} & \minislot{\liquset} & \minislot{\liquset}& \minislot{\liquset}
\end{matrix}
\]
with the additional conditions that the $\tileset{}$ component is a
valid tiling and the $X$ component is made exclusively from the set of
${2\times 2}$ patterns appearing in the following partial
configuration:
\[\small
\begin{matrix}
  &&&  \minislot{\oT} & \minislot{\oT} \\
  &\minislot{\oTL} & \minislot{\oT} & \minislot{\oT} & \minislot{\oT} & \minislot{\oTR} \\
  \minislot{\oL}&\minislot{\oL} & \minislot{\oTL} & \minislot{\oT} & \minislot{\oTR} & \minislot{\oR}& \minislot{\oR} \\
  \minislot{\oL}&\minislot{\oL} & \minislot{\oL} & \minislot{\outo} & \minislot{\oR} & \minislot{\oR}& \minislot{\oR} \\
  &\minislot{\oL} & \minislot{\oBL} & \minislot{\oB} & \minislot{\oBR} & \minislot{\oR}  \\
  &\minislot{\oBL} & \minislot{\oB} & \minislot{\oB} & \minislot{\oB} & \minislot{\oBR}  \\
  &&&\minislot{\oB} & \minislot{\oB} \\
\end{matrix}
\]
The $X$ component is used to give to any cell inside a solid region
a local notion of \emph{inside} and \emph{outside} as depicted by
Figure~\ref{fig:inout} (up to $\pi/2$ rotations): arrows point to
the inside region.

The behaviour of $\acg_{\tileset{}}$ is precisely the following:
\begin{enumerate}
\item if the neighbourhood ($5\times 5$ cells) forms a pattern
  forbidden in $\pobst{\tileset{}}$, then the state is left unchanged
  except for the following cases where it turns into state $0$:
  \begin{itemize}
  \item if the cell is in a liquid state;
  \item if the inside region of the cell forms a forbidden pattern,
  \item the cell together with one of its neighbour forms a forbidden pattern
  \end{itemize}    
\item else, apply (if possible) one of the transition rules depending
  only on the ${3\times 3}$ neighbourhood detailed in
  Figure~\ref{fig:defi} (replacing $\obstset$ by $\nobstset{\tileset{}}$);
\item in any other case, leave the state unchanged if it is solid
  and turn into $0$ if it is liquid.
\end{enumerate}

As for $\acf$, a configuration is said \emph{finite} if it contains
only a finite number of cells in a solid state.

\begin{lemma}[erosion result]
  \label{lem:erodedbis}
  Let $\tileset{}$ be any tile set.  Let ${x\in\pobst{\tileset{}}}$ be
  a finite configuration. Then the set ${X = \{z\in\zz :
    x(z)\in\nobstset{\tileset{}}\}}$ is a union of disjoint squares
  with sides of odd length containing the state '$\outo$' at the
  center, and which are pairwise spaced by at least 2 cells.
\end{lemma}
\begin{proof}
  By definition of $\pobst{\tileset{}}$, $x$ is necessarily made of
  rectangular obstacles which are pairwise spaced by at least $2$
  cells.

  Moreover, the $X$ component ensures that the border of any
  rectangular obstacle is made as follows:
  \begin{itemize}
  \item only state $\oL$ (resp. $\oR$, $\oT$ and $\oB$) on the left
    (resp. right, top and bottom) side;
  \item only state $\oTL$ (resp. $\oTR$, $\oBR$ and $\oBL$) on the
    top-left (resp. top-right, bottom-right and bottom-left) corner.
  \end{itemize}
  Remark that the $X$ component requires that the sequence of state
  obtained by starting from a corner and advancing in the
  corresponding diagonal direction is a succession of identical
  diagonal arrows, then the state '$\outo$' and then a sequence of
  opposite diagonal arrows. This implies that the obstacle is a square
  of odd side length and that the state '$\outo$' is in the
  center.\qed
\end{proof}

From now on, we call \emph{valid obstacle} for $\acg_{\tileset{}}$ a
${n\times n}$ square ($n$ odd) of solid states with state '$\outo$' in
the center and forming a valid pattern of $\pobst{\tileset{}}$.

\begin{lemma}[conservative erosion process]
  \label{lem:finiteattrakbis}
  Let $\tileset{}$ be any tile set.  For any finite configuration $x$
  we have the following:
  \begin{enumerate}
  \item there exists $t_0$ such that, ${\forall t\geq t_0}$,
    ${\acg_{\tileset{}}^t(x)\in\pobst{\tileset{}}}$ and, in
    ${\acg_{\tileset{}}^t(x)}$, any occurrence of $U$ or $D$ is on the
    left of any occurrence of any state from $\nobstset{\tileset{}}$;
  \item if $z_0$ and ${n\geq 7}$, $n$ odd, are such that $x$ contains
    a valid ${n\times n}$ obstacle centered on $z_0$ then ${\forall
      t\geq 0}$ ${\acg_{\tileset{}}^t(x)}$ contains the same valid
    ${(n-4)\times(n-4)}$ square obstacle centered on $z_0$
  \end{enumerate}
\end{lemma}
\begin{proof}
  The first part of this lemma follows by applying arguments of the
  proof of Lemma~\ref{lem:finiteattrak} to $\acg_{\tileset{}}$. The
  only point to check is that given any forbidden pattern for
  $\pobst{\tileset{}}$ we have (straightforward from the definition of
  $\pobst{\tileset{}}$ and interior regions):
  \begin{itemize}
  \item either a pair of cells at distance at most $2$, both in a
    solid state, and which form a forbidden pattern by themselves,
  \item or a cell in a solid state whose inside region forms a
    forbidden pattern.
  \end{itemize}
  Thus, the number of cells in a solid state is guaranteed to decrease
  while the current configuration is not in
  $\pobst{\tileset{}}$. Therefore $\pobst{\tileset{}}$ is reached in
  finite time (any configuration without solid states belongs to
  $\pobst{\tileset{}}$).

  For the second part of the lemma, consider all cells $z$ of the
  lattice such that ${\|z - z_0\|_\infty \leq \frac{n-5}{2}}$
  (i.e. cells belonging to the ${(n-4)\times (n-4)}$ square centered
  on $z_0$). Initially, those cells have a valid neighbourhood so
  after one step, they all stay in the same state. Therefore, by
  definition of a valid square obstacle, they all have a valid
  interior region after one step. Moreover, in their exterior region,
  they all have either valid solid states as in the initial step, or
  liquid states (if some cell at the boundary of the ${n\times n}$
  square has turned into state $0$): in any case, by definition of
  exterior regions, no such cell $z$ has a cell in its neighbourhood
  to form a forbidden pattern with. Therefore, all cells $z$ stay
  unchanged after two step, and the reasoning can be iterated
  forever.\qed
\end{proof}

The infiltration lemma (Lemma~\ref{lem:partattack} for $\acf$) remains
true here, simply replacing $\obst$ by $\pobst{\tileset{}}$. Combined
with the above lemmas, it implies the following proposition.

\begin{proposition}
  \label{prop:onioncarak}
  Let $\tileset{}$ be any tile set. Then $\acg_{\tileset{}}$ is
  sensitive to initial conditions if $\tileset{}$ does not tile the
  plane, and it admits equicontinuous points if $\tileset{}$ tiles the
  plane. Moreover, in the latter case, any equicontinuous point has
  the following properties:
  \begin{itemize}
  \item it is made only of solid states;
  \item it contains exactly one occurrence of state '$\outo$';
  \item its $\tileset{}$ component forms a valid tiling.
  \end{itemize}
\end{proposition}
\begin{proof}
  First, suppose that $\tileset{}$ cannot tile the plane. Then there
  exists $n$ such that there is no valid square tiling of size
  ${n\times n}$. Using the same reasoning as in
  Proposition~\ref{prop:tileac}, we deduce that $\acg_{\tileset{}}$ is
  sensitive to initial conditions (because, by
  Lemma~\ref{lem:finiteattrakbis}, after some time a liquid state must
  appear at some position $z$ with ${\|z\|_\infty\leq n}$ and the
  infiltration can be applied to that position).

  Now suppose that $\tileset{}$ can tile the plane. Consider the
  configuration $x$ made only of solid states and such that:
  \begin{itemize}
  \item the $\tileset{}$ component is a valid tiling;
  \item the $X$ component is made of state $\outo$ is at position
    $(0,0)$ and completed everywhere in a valid way.
  \end{itemize}
  Since any ${n\times n}$ square centered on position $(0,0)$ is a
  valid square obstacle, Lemma~\ref{lem:finiteattrakbis} shows that
  $x$ is an equicontinuous point. Indeed, for any $n$ and for any
  configuration $y$ having a valid ${n\times n}$ square obstacle
  centered on position $(0,0)$, we have that the orbits of $x$ and $y$
  under the action of $\acg_{\tileset{}}$ coincide on the central
  ${(n-4)\times(n-4)}$ part.

  Finally, consider any equicontinuous point $x$ of
  $\acg_{\tileset{}}$. Using the reasoning of the first part of this
  proof, we show that $x$ contains only solid states and that its
  $\tileset{}$ component forms a valid tiling. Moreover, suppose that
  the $X$ component contain at least $2$ occurrences of state
  '$\outo$' and let $n$ be such that $2$ occurrences of '$\outo$' are
  contained in the ${n\times n}$ central square of $x$. By
  Lemmas~\ref{lem:finiteattrakbis} and~\ref{lem:erodedbis}, for any
  finite configuration $y$ identical to $x$ on the central ${n\times
    n}$ region, there is some time after which some cell in the
  central ${n\times n}$ region is in a liquid state (because no valid
  obstacle can contain two occurrences of '$\outo$'). From that point,
  the infiltration argument can be applied, contradicting the fact
  that $x$ is an equicontinuous point. To conclude the proposition, it
  remains the case where the configuration $x$ considered contains no
  occurrence of '$\outo$'. This case is treated as above, since valid
  square obstacles must contain an occurrence of '$\outo$' as stated
  by Lemma~\ref{lem:erodedbis}.\qed
\end{proof}

\subsection{Combining two solid components}
\label{sec:combo}

Our last variation, called $\ach_{\tileset{}}$, is a simple
combination of $\acf$ and $\acg_{\tileset{}}$ (for any given tile set
$\tileset{}$).  More precisely, it is the CA defined over state set
${\obstset\cup\nobstset{\tileset{}}\cup\liquset}$ with the following
behaviour:
\begin{itemize}
\item if the neighbour contains only states from ${\obstset\cup\liquset}$ then behave like $\acf$;
\item if the neighbour contains only states from
  ${\nobstset{\tileset{}}\cup\liquset}$ then behave like $\acg_{\tileset{}}$;
\item in any other case, turn into state $0$.
\end{itemize}

Using what was previously established for $\acf$ and
$\acg_{\tileset{}}$, we have the following proposition for
$\ach_{\tileset{}}$.

\begin{proposition}
  \label{prop:combocarak}
  Let $\tileset{}$ be any tile set. Then $\ach_{\tileset{}}$ is not
  sensitive to initial conditions and it admits equicontinuous points
  if and only if $\tileset{}$ tiles the plane.
\end{proposition}
\begin{proof}
  Since arbitrarily large obstacles of type $\obstset$ can be formed,
  the reasoning of the proof of Proposition~\ref{prop:nosens} can be
  applied here showing that $\ach_{\tileset{}}$ is not sensitive to
  initial conditions.

  Moreover, any equicontinuous point of $x$ of $\acg_{\tileset{}}$ is
  an equicontinuous point of $\ach_{\tileset{}}$. Indeed, for any $n$,
  any configuration $y$ identical to $x$ is the central ${n\times n}$
  region verifies that at any time $t$, the central ${n\times n}$
  region of ${\ach_{\tileset{}}^t(y)}$ is made only of states from
  ${\nobstset{\tileset{}}\cup\liquset}$ and is therefore governed by
  $\acg_{\tileset{}}$. Thus, the reasoning of
  Proposition~\ref{prop:onioncarak} applies here. Hence, if
  $\tileset{}$ can tile the plane, then $\ach_{\tileset{}}$ admits
  equicontinuous points.

  Conversely, suppose that $\tileset{}$ cannot tile the plane. So
  $\ach_{\tileset{}}$ has no equicontinuous points in
  ${(\nobstset{\tileset{}}\cup\liquset)^\zz}$ because it would be an
  equicontinuous point for $\acg_{\tileset{}}$, thus contradicting
  Proposition~\ref{prop:onioncarak}. Similarly, there cannot be
  equicontinuous point in ${(\obstset\cup\liquset)^\zz}$ because it
  would contradict Proposition~\ref{prop:noequ}. Finally, a
  configuration $x$ containing states from both sets $\obstset$ and
  $\nobstset{\tileset{}}$ cannot be an equicontinuous point either
  because $x$ or $\ach_{\tileset{}}(x)$ necessarily contains a liquid
  state and in such a case the infiltration argument can be applied as
  in Proposition~\ref{prop:noequ} (Lemmas~\ref{lem:eroded},
  \ref{lem:finiteattrak} and \ref{lem:partattack} are true for
  $\ach_{\tileset{}}$).\qed
\end{proof}

\begin{figure}
  \centering
  \begin{tabular*}{0.8\textwidth}{@{\extracolsep{\fill}} c|l|l}
    \textit{Automaton} & \textit{Solid component} & \textit{Behaviour}\\
    \hline
    &&\\
    $\acf$ & $\obstset$ & $\nono$\\
    $\acf_{\tileset{}}$ & $\obstset_{\tileset{}}$ & $\sensi$ if $\tau$ tiles, $\nono$ else\\
    $\acg_{\tileset{}}$ & $\nobstset{\tileset{}}$ & $\equpt$ if $\tau$ tiles, $\sensi$ else\\
    $\ach_{\tileset{}}$ & $\nobstset{\tileset{}}\cup\obstset$ & $\equpt$ if $\tau$ tiles, $\nono$ else\\
  \end{tabular*}
  \caption{Summary of constructions}
  \label{fig:summary}
\end{figure}

\section{Topological Classification Revisited}
\label{sec:classif}

Equipped with the various constructions detailed above (see
Figure~\ref{fig:summary}), we study in this section the topological
classification of P.~K\r urka (put aside expansivity) for higher
dimensional cellular automata.

In \cite{varouch}, the authors give a recursive construction which
produce either a 1D CA with equicontinuous points or a 1D sensitive CA
according to whether a Turing machine halts on the empty input or
not. By Proposition~\ref{prop:dimraise}, we get the following result.

\begin{proposition}
  \label{prop:varouch}
  For any dimension, the classes $\sensi$ and $\equpt$ are recursively
  inseparable. Moreover, $\sensi$ is not recursively enumerable and
  $\equpt$ is not co-recursively enumerable.
\end{proposition}

However, this is not enough to establish the overall undecidability of
the topological classification of 2D CA. The main concern of this
section is to complete Proposition~\ref{prop:varouch} in order to
prove a stronger and more complete undecidability result summarised in
the following theorem.

\begin{theorem}
  \label{theo:undeci}
  For any dimension strictly greater than $1$, we have the following:
  \begin{itemize}
  \item each of the classes $\equpt$, $\sensi$ and $\nono$ is neither
    recursively enumerable nor co-recursively enumerable;
  \item any pair of them is recursively inseparable.
  \end{itemize}
\end{theorem}
\begin{proof}
  The proof of this theorem is made of $3$ similar parts: each one
  gives the inseparability of two classes $A$ and $B$ among $\sensi$,
  $\equpt$ and $\nono$, as well as the non enumerability of $A$ and
  the non co-enumerability of $B$.  The propositions focus on 2D
  cellular automata but, by Proposition~\ref{prop:dimraise}, results
  remain true for higher dimensions (because the canonical lift from
  some CA $\acf$ to $\raisdim{\acf}$ is recursive).  The $3$ parts are
  proved in the following way:
  \begin{description}
  \item[$A=\sensi$ and $B=\equpt$: ]this is
    Proposition~\ref{prop:varouch} (our construction
    $\acg_{\tileset{}}$ gives an alternative proof by Berger's
    theorem).
  \item[$A=\nono$ and $B=\sensi$: ]this follows by Berger's
    theorem~\cite{berger} (the set of tile sets which can tile the
    plane is not recursively enumerable) and
    Proposition~\ref{prop:tileac} since $\acf_{\tileset{}}$ can be
    recursively constructed from $\tileset{}$.
  \item[$A=\equpt$ and $B=\nono$: ]again since the set of tile sets
    that can tile the plane is not recursively enumerable, this
    follows by Proposition~\ref{prop:combocarak}.\qed
  \end{description}
\end{proof}

Besides complexity of decision problems, other differences appears
between dimension 1 and higher dimensions. Let us first stress the
dynamical consequence of the construction of CA $\acf_{\tileset{}}$.
It is well-known that for any 1D sensitive CA of radius $r$,
${2^{-2r}}$ is always the maximal admissible sensitivity constant (see
for instance~\cite{Kurka97}). Thanks to the above construction it is
easy to construct CA with tiny sensitivity constants as shown by the
following proposition.

\begin{proposition}
  \label{prop:constant}
  The (maximal admissible) sensitivity constant of sensitive 2D CA
  cannot be recursively (lower-)bounded in the number of states and
  the neighbourhood size.
\end{proposition}
\begin{proof}
  This follows directly from Proposition~\ref{prop:tileac} since the
  size $n$ of the largest ${n\times n}$ valid tiling for a given tile
  set is not a recursive function of the tile set.\qed
\end{proof}

To finish this section, we will discuss another difference between 1D
and 2D concerning the complexity of equicontinuous points. Let us
first recall that equicontinuous point in 1D CA can be generated by
finite words often called ``blocking'' words. A finite word $u$ is
blocking for some CA $\acf$ if for any pair of configurations $x$ and
$y$ both having pattern $u$ in their center, we have\footnote{To
  simplify the definition, we require that the blocking word fixes the
  ${2r+1}$ central columns of the space-time diagrams of any
  configuration having $u$ in its center. In fact $2r$ columns would
  be enough (and it is the standard definition) but it doesn't change
  anything for our purpose since with our definition of blocking word,
  we still have the property that a 1D CA admits equicontinuous points
  if and only if it has a blocking word.}:
\[\forall t\geq 0, \forall z: \|z\|_\infty\leq r\Rightarrow \acf^t(x)(z)=\acf^t(y)(z)\]
where $r$ is the radius of $\acf$.

For any $\acf$ with equicontinuous points, there exists a finite word
$u$ such that ${{}^\infty u^\infty}$ is an equicontinuous point for
$\acf$ (proof in~\cite{Kurka97}). The construction $\acg_{\tileset{}}$
can be used with the tile set of Myers~\cite{myers} which can produce
only non-recursive tilings of the plane. Therefore the situation is
more complex in 2D, and we have the following proposition.

\begin{proposition}
  \label{prop:nonrecpoint}
  For any dimension strictly greater than $1$, there exists a CA
  having equicontinuous points, but only non-recursive ones.
\end{proposition}
\begin{proof}
  By Proposition~\ref{prop:onioncarak}, any equicontinuous point of
  $\acg_{\tileset{}}$ is made solely of solid states and its
  $\tileset{}$ component forms a valid tiling. Now consider the tile
  set $\tileset{0}$ of Myers~\cite{myers}: it can tile the plane but
  only with non-recursive tilings. Therefore, by
  Proposition~\ref{prop:onioncarak}, $\acg_{\tileset{0}}$ admits
  equicontinuous points, but only non recursive ones\qed
\end{proof}

\begin{remark}
  Since the construction $\acg_{\tileset{}}$ enforces the apparition
  of a particular state ($\outo$) in any equicontinuous point, we
  could have proved Proposition~\ref{prop:nonrecpoint} using the
  simpler tile set of Hanf \cite{hanf}, which produces only
  non-recursive tilings provided some fixed tile is placed at the
  origin.
\end{remark}

Any 1D CA with equicontinuous points, admits in fact uncountably many
equicontinuous points. Indeed, if $u$ is a blocking word and if $c$ is
any bi-infinite sequence of $0$ or $1$, then the configuration:
\[\cdots c(-n)\cdot u\cdots c(-1)\cdot u\cdot c(0)\cdot u\cdots c(n)\cdots\]
is always an equicontinuous point.  The next proposition shows that it
is no longer the case for higher dimensional CA.

\begin{proposition}
  \label{prop:countequpt}
  For any dimension strictly greater than $1$, there exists a CA
  having a countably infinite set of equicontinuous points.
\end{proposition}
\begin{proof}
  Let $\tileset{0}$ be a trivial tile set (a single tile and no
  constraint). By Proposition~\ref{prop:onioncarak},
  $\acg_{\tileset{0}}$ admits equicontinuous points which are all
  identical on their tiling component. Moreover, it follows from
  definition of $\pobst{\tileset{0}}$ that if two equicontinuous
  points have the state '$\outo$' in the same position, then they are
  identical. Thus $\acg_{\tileset{0}}$ possesses only a countable set
  of equicontinuous points and the proposition follows for dimension
  $2$.

  For dimension $3$ we will use a lifted version $\acg_+$ of
  $\acg_{\tileset{0}}$: $\acg_+$ is essentially a canonical lift of
  $\acg_{\tileset{0}}$ with the additional condition that $2$ cells
  whose coordinates differ by $1$ only on the third dimension must be
  in the same state, otherwise they turn into state $0$ whatever the
  2D dynamics of $\acg_{\tileset{0}}$ says. By a straightforward
  adaptation of the reasoning of Proposition~\ref{prop:onioncarak} we
  have the following: for any equicontinuous point of $\acg_+$, the
  set of occurrences of states '$\outo$' is exactly a line co-linear to
  the third dimension. Therefore, by the same reasoning as above, we
  deduce that $\acg_+$ has only a countable set of equicontinuous
  points.
  
  The lift arguments can be iterated and thus the proposition follows
  for any dimension.\qed
\end{proof}

\section{Complexity of Sensitivity According to Dimension}
\label{sec:arithier}

In this section, we study the complexity of the set of $\sensi$ from
the point of view of the arithmetical hierarchy. More precisely, we
establish an upper bound in the 1D case and a lower bound in the 3D
case showing that the complexity of $\sensi$ does vary with dimension.

\begin{proposition}
  \label{prop:sens1D}
  For 1D cellular automata, the set $\sensi$ is $\Pi_2^0$.
\end{proposition}
\begin{proof}
  As said above, a 1D CA is sensitive if and only if it does not
  possess any blocking word \cite{Kurka97}. Let $\acf$ be a CA of
  radius $r$. Following the definition of blocking words given in
  Section~\ref{sec:classif}, the fact that $\acf$ possesses a blocking word
  can be expressed as follows:
  \[\exists u\, \forall t\ R(u,t)\]
  where $R(u,t)$ is true if and only if for all $t'\leq t$ and all
  pair of configurations $x$ and $y$ having $u$ in their center, we have:
  \[\forall z: \|z\|_\infty\leq r\Rightarrow \acf^{t'}(x)(z)=\acf^{t'}(y)(z).\]
  $R(u,t)$ is recursive since the checking involve only a finite part
  of the initial configuration (precisely the ${2r(t+1)}$ central
  cells).  Hence, the set $\sensi$ is characterised by the $\Pi_2^0$
  predicate ${\forall u\,\exists t\ \neg R(u,t)}$.\qed
\end{proof}

We will now give a hardness result for the set $\sensi$ in dimension
$3$. We will reduce \cofin, the set of Turing machines halting on a
co-finite set of inputs, to $\sensi$ thus proving that $\sensi$ is
$\Sigma_3^0$-hard (see~\cite{rogers} for the proof of
$\Sigma_3^0$-completeness of \cofin).

We will use simulations of Turing machines by tile sets in the
classical way (originally suggested by Wang~\cite{Wang}): the tiling
represents the space-time diagram of the computation and the
transition rule of the Turing machine are converted into tiling
constraints.  For technical reasons which will appear clearly in the
proof of Lemma~\ref{lem:reonion}, we slow down the computation (what
can be done by a recursive modification of the machine): the head
takes $2$ time steps to move $1$ cell left or right.  Moreover, the
tile sets we consider always contain some blank tile $\beta$
(corresponding to a blank tape symbol of the Turing machine) and some
special tile $\alpha$ used to initiate the computation, but \emph{no
  tile corresponding to a final state of the Turing machine}. More
precisely, each tile set enforces the following:
\begin{itemize}
\item if some row contains $\alpha$, it is of the form
  ${}^\infty\beta\alpha w\beta^\infty$ where $w$ is a sequence of
  non-blank symbols which will be treated as input (at this point we
  can not enforce by tiling constraint that $w$ is finite);
\item the tile on the right of $\alpha$ must represent the Turing head
  in its initial state reading the first letter of the input.
\end{itemize}
Thus, each time a valid tiling contains $\alpha$, we are guaranteed
that it contains a valid non-halting computation starting on some
(potentially infinite) input.

The $i^{th}$ Turing machine in a standard enumeration is denoted by
$\machine{i}$ and to each $\machine{i}$ we associate a tile set
$\tileset{i}$ whose constraints ensure the simulation of $\machine{i}$
as mentioned above, and which contains the special tiles $\alpha_i$
and $\beta_i$ as described above.

We now describe the construction, for any Turing machine
$\machine{i}$, of a cellular automaton $\aci_i$ which is sensitive to
initial conditions if and only if ${\machine{i}\in\cofin}$. It will
essentially consist in a lift to dimension $3$ of a modified version
of $\acg_{\tileset{i}}$. We first describe this modified version,
denoted $\acgp{i}$, which is a 2D CA.

The intuition is the following: we want that any equicontinuous point
of $\acgp{i}$ contains a valid non-halting computation of
$\machine{i}$ starting from a finite input. More precisely, we will
define $\acgp{i}$ in such a way that any equicontinuous point has a
valid $\tileset{i}$-tiling on some of its components, which contains
an occurrence of the special state $\alpha_i$, and which contains only a
\emph{finite} sequence of non blank symbols on the right of
$\alpha_i$.

The definition of $\acgp{i}$ differ from that of $\acg_{\tileset{i}}$
only by the definition of the subshift $\pobst{\tileset{i}}$: for
$\acgp{i}$ this subshift becomes $\ppobst{i}$ defined as follows.
A configuration $x$ is in $\ppobst{i}$ exactly when:
\begin{itemize}
\item $x\in\pobst{\tileset{i}}$;
\item $\alpha_i$ is the only tile allowed in the tiling component of
  a state having its $X$ component equal to $\outo$;
\item a solid state having a tile different from $\beta_i$ in its
  tiling component is not allowed to be on the immediate left of a
  liquid state.
\end{itemize}

$\acgp{i}$ is built upon $\ppobst{i}$ exactly as $\acg_{\tileset{i}}$
is built upon $\pobst{\tileset{i}}$. Precisely, any cell of $\acgp{i}$
behave like this:

\begin{enumerate}
\item if the neighbourhood ($5\times 5$ cells) forms a pattern
  forbidden in $\ppobst{i}$, then the state is left
  unchanged except in the following cases where it turns into state
  $0$:
  \begin{itemize}
  \item if the cell is in a liquid state;
  \item if the inside region of the cell forms a forbidden pattern,
  \item the cell together with one of its neighbour forms a forbidden pattern
  \end{itemize}    
\item else, apply (if possible) one of the transition rules depending
  only on the ${3\times 3}$ neighbourhood detailed in
  Figure~\ref{fig:defi} (replacing $\obstset$ by $\nobstset{\tileset{i}}$);
\item in any other case, leave the state unchanged if it is solid
  and turn into $0$ if it is liquid.
\end{enumerate}

From this definition and the result already established for
$\acg_{\tileset{i}}$ we easily get the following lemma.

\begin{lemma}
  \label{lem:reonion}
  $\acgp{i}$ is sensitive to initial conditions if $\machine{i}$ halts
  on any input. Moreover, if $\machine{i}$ doesn't halt on all inputs,
  then $\acgp{i}$ admits equicontinuous points and each equicontinuous
  point verifies the following:
  \begin{itemize}
  \item its tiling component forms a valid tiling for $\tau_i$;
  \item it contains exactly one occurrence of the special tile $\alpha_i$;
  \item there is a finite sequence $w$ of consecutive non-blank
    symbols on the right of $\alpha_i$, therefore the tiling component
    simulates a valid non-halting computation of $\machine{i}$
    starting on a finite input $w$.
  \end{itemize}
\end{lemma}
\begin{proof}
  The modifications introduced in $\acgp{i}$ (compared to
  $\acg_{\tileset{i}}$) concern only new cases in which a solid state
  is turned into $0$. Therefore, all necessary conditions about
  equicontinuous points of $\acg_{\tileset{i}}$
  (Proposition~\ref{prop:onioncarak}) apply here.  Besides, if
  $\machine{i}$ possesses a non-halting input, it is easy to construct
  an equicontinuous point $x$ which contains a valid space-time
  diagram of a non-halting computation. The fact that the computation
  is slow ensures that that we can find arbitrarily large squares
  centered on the tile $\alpha_i$ (and the state $\outo$) without any
  non-blank on the right boundary of the square. With such
  precautions, the conservative erosion apply here exactly as in the
  proof of Proposition~\ref{prop:onioncarak}.

  Finally, since the definition of $\acgp{i}$ implies that occurrences
  of $\outo$ coincide with occurrences of $\alpha_i$, the lemma
  follows from the following property: if a configuration $x$ of
  $\acgp{i}$ contains a cell having an infinite sequence of non-blank
  symbols on its right, then it is not an equicontinuous point. This
  property follows from the definition of $\ppobst{i}$ since, for any
  finite configuration sufficiently close to $x$, the non-blank
  symbols allow liquid states to infiltrate towards a fixed position
  (after some time) and therefore the usual technique of particle
  infiltration shows that $x$ cannot be an equicontinuous point.\qed
\end{proof}

\paragraph{The $3$-dimensional cellular automaton $\aci_i$.} The idea
is that on each horizontal plane ${\plane{c} = \{(a,b,c) :
  a,b\in\zz\}}$ of the space, $\aci_i$ generally behaves like
$\acgp{i}$. However, $\aci_i$ contains an additional 3D mechanism,
whose role is to ensure that the non-halting simulations done on
successive planes start from different inputs of
$\machine{i}$. $\aci_i$ contains an additional component of states,
called $Z$, that can take $3$ values '$\zplus$', '$\zminus$' and
'$\zzero$' (the state set of $\aci_i$ is ${Q_i\times Z}$ where $Q_i$
is the state set of $\acgp{i}$). To describe the local constraints on
$Z$, we use notations $\sud{\cdot}$, $\nor{\cdot}$, $\est{\cdot}$,
$\oue{.}$ to describe relation between positions in the same
horizontal plane, and $\hau{\cdot}$ and $\bas{\cdot}$ for the
$3^\text{rd}$ dimension:
\begin{itemize}
\item if the $Z$-component of a cell $z\in\Z^3$ is '$\zzero$' then it
  is also the case for cells $\est{z}$, $\oue{z}$,
  $\hau{z}$ and $\bas{z}$;
\item if the $Z$-component of a cell $z\in\Z^3$ is '$\zplus$' then it
  is also the case for cells $\est{z}$, $\oue{z}$ and $\hau{z}$, whereas
  $\nor{z}$ and $\sud{z}$ must have a $Z$-component equal to
  '$\zzero$';
\item if the $Z$-component of a cell $z\in\Z^3$ is '$\zminus$' then it
  is also the case for cells $\est{z}$, $\oue{z}$ and $\bas{z}$, whereas
  $\nor{z}$ and $\sud{z}$ must have a $Z$-component equal to
  '$\zzero$';
\item if the tiling component of a cell $z$ (in a solid state) is
  $\alpha_i$ then its $Z$-component must be either '$\zplus$' or
  '$\zminus$'; moreover $\hau{z}$ and $\bas{z}$ must also be in a
  solid state with a tiling component equal to '$\alpha_i$';
\item if a cell $z$ in a solid state has its $Z$-component equal to
  '$\zplus$' and its tiling component is $\beta_i$, then, if $\oue{\bas{z}}$
  has also its $Z$-component equal to '$\zplus$' and is also solid, it
  must have its tiling component also equal to $\beta_i$;
\item if a cell $z$ in a solid state has its $Z$-component equal to
  '$\zminus$' and its tiling component is $\beta_i$, then, if $\oue{\hau{z}}$
  has also its $Z$-component equal to '$\zplus$' and is also solid, it
  must have its tiling component also equal to $\beta_i$.
\end{itemize}

The global result of those local conditions is illustrated by the
following lemma.

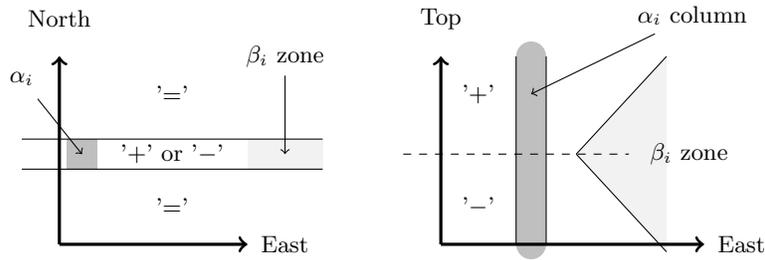
\begin{figure}[htbp]
  \hbox to\textwidth{%
    \hfill%
    \begin{tikzpicture}
      \fill[fill=gray!10] (2.5,1) -- (2.5,1.4) -- (3.5,1.4) -- (3.5,1);
      \fill[fill=gray!50] (.1,1.4) -- (.5,1.4) -- (.5,1) -- (.1,1);
      \draw (-.5,1.4) -- (3.5,1.4);
      \draw (-.5,1) -- (3.5,1);
      \draw [very thick, ->] (0,0) -- (0,2.5);
      \path (0,3) node () {North};
      \draw [very thick, ->] (0,0) -- (2.5,0);
      \path (3,0) node () {East};
      \path (1.5,2) node () {'$=$'};
      \path (1.5,.5) node () {'$=$'};
      \path (1.5,1.2) node () {'$+$' or '$-$'};
      \path (-.5,2.2) node[shape=circle] (apos) {$\alpha_i$};
      \draw[->] (apos) -- (.3,1.2) node () {};
      \draw (3,2.5) node[] (bpos) {$\beta_i$ zone};
      \draw[->] (bpos) -- (3,1.2) node () {};
    \end{tikzpicture}\hfill%
    \begin{tikzpicture}
      \fill[fill=gray!10] (1.8,1.2)--(3,2.5)--(3,-.1);
      \draw (1.8,1.2)--(3,2.5);
      \draw (1.8,1.2)--(3,-.1);
      \draw (3.3,1.2) node () {$\beta_i$ zone};
      \draw[line width=0.4cm,color=gray!50,cap=round] (1.2, 0) -- (1.2,2.5);
      \draw (1.4,0) -- (1.4,2.5);
      \draw (1,0) -- (1,2.5);
      \draw[dashed] (-.5,1.2) -- (2.5,1.2);
      \path (.5,2) node () {'$+$'};
      \path (.5,.5) node () {'$-$'};
      \draw [very thick, ->] (0,0) -- (0,2.5);
      \path (0,3) node () {Top};
      \draw [very thick, ->] (0,0) -- (3.5,0);
      \path (4,0) node () {East};
      \path (2.5,2.8) node[anchor=south west] (pos) {$\alpha_i$ column};
      \draw[->] (pos) -- (1.2,2) node () {};
    \end{tikzpicture}\hfill%
  }
  \caption{Two planar (simplified) views of a valid solid
    configuration.}
  \label{fig:zcomponent}
\end{figure}

\begin{lemma}
  \label{lem:alignage}
  Let $x$ be a purely solid configuration of $\aci_i$ such that, each
  horizontal plane contains one occurrence of $\alpha_i$ and a valid
  tiling, and all the previous local conditions are verified. Then $x$
  has the following form:
  \begin{itemize}
  \item on each plane, all $Z$ components are '$\zzero'$ except on an
    east/west line which contains $\alpha_i$;
  \item all the occurrences of $\alpha_i$ are aligned in a top/bottom
    column;
  \item the space is made of a top half corresponding to planes
    $\plane{c}$ having some state with $Z$-component '$\zplus$' and a
    bottom half corresponding to planes $\plane{c}$ having some state
    with $Z$-component '$\zminus$';
  \item if a plane $\plane{c}$ is in the top half and simulates
    $\machine{i}$ on an input of length $n$, then for any ${a>0}$, the
    plane $\plane{c+a}$ simulates $\machine{i}$ on an input of length
    strictly greater than $n$;
  \item similarly for the bottom half, the input length is strictly
    greater for plane $\plane{c-a}$ than for plane $\plane{c}$.
  \end{itemize}
\end{lemma}
\begin{proof}
  Straightforward.\qed
\end{proof}

$\aci_i$ is then defined as follows: if one of the previous local
conditions is violated in the neighbourhood of a cell in a solid state
surrounded only by cells in a solid state, then the cell turns into
state $(0,\zzero)$, else it behaves according to $\acgp{i}$ depending
only on cells in the same plane.

\begin{proposition}
  For dimension $3$, the set $\sensi$ is $\Sigma_3^0$-hard.
\end{proposition}
\begin{proof}
  We show that $\aci_i$ is sensitive to initial conditions if and only
  if $\machine{i}$ admits an infinite set of non-halting inputs,
  which yields a reduction from \cofin{} to $\sensi$.

  First, it is easy to see that if $\machine{i}$ has an infinite set
  of non-halting inputs, then an equicontinuous point for $\aci_i$ can
  be build: given an infinite sequence of non-halting inputs of
  different lengths, one can build a purely solid configuration, made
  of two halves, each one corresponding to the sequence of valid
  simulations on each plane for successive inputs, and respecting all
  the conditions on the $Z$ component. It is straightforward to check
  that such a configuration is an equicontinuous point. 

  Conversely, if $x$ is an equicontinuous point for $\aci_i$ then each
  plane $\plane{c}$ must be an equicontinuous point for $\acgp{i}$
  when we forget the $Z$ component. Indeed, the additional 3D
  conditions of $\aci_i$ never affect liquid states and can only turn
  a solid state into state $0$. Now, adding 3D constraints, we deduce
  by Lemmas~\ref{lem:reonion} and \ref{lem:alignage} that
  $\machine{i}$ must have an infinite set of non-halting inputs.\qed
\end{proof}

\newpage
\section{Future Work}
\label{sec:open}


In this paper, we adopted the classical framework of topological
dynamics (which does not explicitly refer to dimension) and studied
how its application to cellular automata may vary with dimension.

The first research direction opened by this paper is the study of new
dynamical behaviour appearing in dimension 2 and more. Indeed, the
mechanisms of information propagation can no longer be explained by
the presence of particular finite words (blocking words in dimension
1). In this general direction, the following questions seems
particularly relevant to us:
\begin{itemize}
\item what kind of dynamics can be found in the class $\nono$?
\item what kind of 2D cellular automata can be built which are in
  $\equpt$ and have a set of equicontinuous points of full measure?
  can we characterise such CA?
\item what happens when we restrict to reversible cellular automata?
  more generally to surjective ones?
\end{itemize}

The second part of the paper concerns complexity of decision problems
related to topological dynamics properties. Our construction
techniques allow to prove several complexity lower bounds. However,
upper bounds seems harder to establish. We think the following
questions are worth being investigated:

\begin{itemize}
\item what is the exact complexity of $\sensi$ in 1D? is it
  $\Pi_2$-complete or only at level $1$ of the arithmetical hierarchy?
\item we believe that the set $\sensi$ is in the arithmetical
  hierarchy for any dimension, but we have no proof yet starting from
  dimension $2$.
\item can we generally implement ``Turing-jumps'' in the complexity of
  the problem we consider when we increase dimension? or is there
  limitation coming from the nature of the problem?
\end{itemize}

Finally, the various kind of sensitivity to dimension change we
encountered, suggest to consider those problems from of more general
point of view by allowing the lattice of cells to be any Cayley graph.
Can we then characterise graphs for which $\sensi$ and $\equpt$ are
complementary classes? What can be said on the complexity of the
different classes of topological dynamics?

\bibliographystyle{splncs}
\bibliography{ac}

\begin{thebibliography}{10}

\bibitem{cie}
Sablik, M., Theyssier, G.:
\newblock Topological dynamics of 2d cellular automata.
\newblock In: CiE. (2008)  523--532

\bibitem{Kurka97}
K\r{u}rka, P.:
\newblock Languages, equicontinuity and attractors in cellular automata.
\newblock Ergodic Theory and Dynamical Systems \textbf{17} (1997)  417--433

\bibitem{BlanchardMaass}
Blanchard, F., Maass, A.:
\newblock Dynamical properties of expansive one-sided cellular automata.
\newblock Israel J. Math. \textbf{99} (1997)

\bibitem{tisseur}
Blanchard, F., Tisseur, P.:
\newblock Some properties of cellular automata with equicontinuity points.
\newblock Ann. Inst. Henri Poincar\'e, Probabilit\'es et statistiques
  \textbf{36} (2000)  569--582

\bibitem{permExp}
Fagnani, F., Margara, L.:
\newblock Expansivity, permutivity, and chaos for cellular automata.
\newblock Theory of Computing Systems \textbf{31}(6) (1998)  663--677

\bibitem{varouch}
Durand, B., Formenti, E., Varouchas, G.:
\newblock On undecidability of equicontinuity classification for cellular
  automata.
\newblock In Morvan, M., R{\'e}mila, {\'E}., eds.: DMCS'03. Volume~AB of DMTCS
  Proceedings. (2003)  117--128

\bibitem{kari94}
Kari, J.:
\newblock Reversibility and {S}urjectivity {P}roblems of {C}ellular {A}utomata.
\newblock Journal of Computer and System Sciences \textbf{48}(1) (1994)
  149--182

\bibitem{Bernardi}
Bernardi, V., Durand, B., Formenti, E., Kari, J.:
\newblock A new dimension sensitive property for cellular automata.
\newblock In Fiala, J., Koubek, V., Kratochv{\'i}l, J., eds.: Mathematical
  Foundations of Computer Science 2004. Volume 3153 of Lecture Notes in
  Computer Science., Springer (2004)  416--426

\bibitem{hedlund}
Hedlund, G.A.:
\newblock Endomorphisms and {A}utomorphisms of the {S}hift {D}ynamical
  {S}ystems.
\newblock Mathematical Systems Theory \textbf{3}(4) (1969)  320--375

\bibitem{berger}
Berger, R.:
\newblock The undecidability of the domino problem.
\newblock Mem. Amer. Math Soc. \textbf{66} (1966)

\bibitem{myers}
Myers, D.:
\newblock Nonrecursive tilings of the plane. ii.
\newblock The Journal of Symbolic Logic \textbf{39}(2) (1966)  286--294

\bibitem{hanf}
Hanf, W.:
\newblock Nonrecursive tilings of the plane. i.
\newblock The Journal of Symbolic Logic \textbf{39}(2) (1966)  283--285

\bibitem{rogers}
Rogers, H.:
\newblock Theory of recursive functions and effective computability.
\newblock MIT Press, Cambridge (1967)

\bibitem{Wang}
Wang, H.:
\newblock Proving theorems by pattern recognition ii.
\newblock Bell System Tech. Journal \textbf{40}(2) (1961)

\end{thebibliography}

\end{document}